\documentstyle[12pt]{article}
\begin{document}
\baselineskip=24pt
\def\rd{{\rm d}}
\newcommand{\Lamb}{\Lambda_{b}}
\newcommand{\Lamc}{\Lambda_{c}}
\newcommand{\Lam}{\Lambda}
\newcommand{\dsp}{\displaystyle}
\newcommand{\nn}{\nonumber}
\newcommand{\ra}{\rightarrow}
\newcommand{\dfr}[2]{ \displaystyle\frac{#1}{#2} }
\newcommand{\Lag}{\Lambda \scriptscriptstyle _{ \rm GR} } 
\renewcommand{\baselinestretch}{1.5}
\begin{titlepage}
\vspace{-20ex}
\vspace{1cm}
\begin{flushright}
\vspace{-3.0ex} 
    {\sf ADP-98-15/T291} \\
\vspace{-2.0mm}
\vspace{5.0ex}
\end{flushright}

\centerline{\Large\sf Direct CP Violation in $\Lamb\rightarrow n (\Lam)
\pi^+\pi^-$ Decays via $\rho-\omega$ Mixing}
\vspace{6.4ex}
\centerline{\large\sf  	X.-H. Guo$^{1,2}$ and A.W. Thomas$^{1}$}
\vspace{3.5ex}
\centerline{\sf $^1$ Department of Physics and Mathematical Physics,}
\centerline{\sf and Special Research Center for the Subatomic Structure of
Matter,}
\centerline{\sf University of Adelaide, SA 5005, Australia}
\centerline{\sf $^2$ Institute of High Energy Physics, Academia Sinica,
Beijing 100039, China}
\centerline{\sf e-mail:  xhguo@physics.adelaide.edu.au,
athomas@physics.adelaide.edu.au}
\vspace{6ex}
\begin{center}
\begin{minipage}{5in}
\centerline{\large\sf 	Abstract}
\vspace{1.5ex}
\small {We study direct CP violation in the bottom baryon decays 
$\Lamb\rightarrow f \rho^0 (\omega) \rightarrow f\pi^+\pi^-$ ($f=n$ or $\Lam$).
It is found that in these decays via $\rho-\omega$ mixing
the CP violation could be very large when the invariant
mass of the $\pi^+\pi^-$ pair is in the vicinity of the $\omega$ resonance.
With a typical value $N_c =2$ in the factorization approach, 
the maximum CP-violating asymmetries are more than 50\% and 68\% for 
$\Lamb\rightarrow n \pi^+\pi^-$ and $\Lamb\rightarrow \Lam \pi^+\pi^-$,
respectively. 
With the aid of heavy quark symmetry and phenomenological models
for the hadronic wave functions of $\Lamb$, $\Lam$ and the neutron, we 
estimate the branching ratios of $\Lamb\rightarrow n (\Lam)\rho^{0}$.}

\end{minipage}
\end{center}

\vspace{1cm}

{\bf PACS Numbers}: 11.30.Er, 12.15.Hh, 13.20.He, 12.39.Hg 
\end{titlepage}
\vspace{0.2in}
{\large\bf I. Introduction}
\vspace{0.2in}

CP violation is still an open problem in the Standard Model, even though
it has been known in the neutral kaon system for more than three decades 
\cite{fleischer}. The study of CP violation in other systems 
is important in order to understand whether the Standard Model provides a
correct description of this phenomenon through 
the Cabbibo-Kobayashi-Maskawa (CKM) matrix. 

Recent studies of direct CP violation in the $B$ meson system\cite{carter}
have suggested that large CP-violating
asymmetries should  be observed in forthcoming
experiments. It is also 
interesting to study CP violation in the bottom baryon system in order to 
find the
physical channels which may have large CP asymmetries, even though the 
branching ratios for such processes are usually smaller than  those
for the corresponding processes of bottom mesons. The study of CP
violation in the bottom system will be helpful for understanding the
origin of CP violation and may provide useful information about
the possible baryon asymmetry in our universe. Actually, some data on the 
bottom baryon $\Lamb$ have appeared just recently.
For instance, OPAL has measured its lifetime and the production 
branching ratio for the inclusive semileptonic decay 
$\Lambda_b \rightarrow \Lambda l^- \bar{\nu} X$ \cite{opal}. Furthermore, 
measurements of the nonleptonic decay 
$\Lamb \rightarrow \Lam J/\psi$ have also been reported \cite{cdf}.
More and more data are certainly expected in the future. It is the purpose
of the present paper to study the CP violation problem in the hadronic decays
$\Lamb\rightarrow n \pi^+\pi^-$ and $\Lamb\rightarrow \Lam \pi^+\pi^-$. 

The CP-violating asymmetries in the decays we are considering arise
from the nonzero phase in the CKM matrix, and hence we have the 
so-called direct CP violation which occurs through the interference of
two amplitudes with different weak and strong phases. The
weak phase difference is determined by the CKM matrix elements while
the strong phase is usually difficult to control. 
In Refs.\cite{eno, tony}, the authors studied direct CP violation 
in $B$ hadronic decays 
through the interference of tree and penguin diagrams, where
$\rho-\omega$ mixing was used to obtain a large strong
phase (as required for large CP violation). The data for $e^+ e^-
\rightarrow  \pi^+\pi^-$ in the $\rho-\omega$ interference region strongly
constrains the $\rho-\omega$ mixing parameters.
Gardner {\it et al.} established not only that the
CP-violating asymmetry in $B^{\pm}\rightarrow \rho^{\pm}\rho^0 (\omega)
\rightarrow \rho^{\pm}\pi^+\pi^-$ is more than 20\% when the invariant 
mass of the $\pi^+\pi^-$ pair is near the $\omega$ mass, but that
the measurement of the CP-violating asymmetry for these decays can remove
the mod$(\pi)$ uncertainty in arg$[-V_{td}V^{*}_{tb}/(V_{ud}V^{*}_{ub})]$
\cite{tony}. In the present work we generalize these discussions 
to the bottom baryon case. It will be shown that the CP violation in
$\Lamb$ hadronic decays could be very large.

In our discussions hadronic matrix elements for both tree and penguin
diagrams are involved. These matrix elements are controlled by 
the effects of nonperturbative
QCD which are difficult handle. In order to extract the
strong phases in our discussions we will use the factorization approximation
so that one of the
currents in the nonleptonic decay Hamiltonian is factorized out and
generates a meson. Thus the decay amplitude of the two body nonleptonic decay
becomes the product of two matrix elements, one related to the decay
constant of the factorized meson and the other to the weak transition
matrix element between two hadrons. 

There have been many discussions concerning the
plausibility of the factorization approach. Since bottom hadrons are very 
heavy, their hadronic decays are energetic. Hence the 
quark pair generated by one current in the weak 
Hamiltonian moves very fast away
from the weak interaction point. Therefore, by the time this quark pair
hadronizes into a meson it is far away from other quarks and is therefore
unlikely to interact with the remaining quarks. Hence this quark pair is
factorized out and generates a meson. This argument is based on the idea
of ``color transparency'' proposed by Bjorken \cite{bjorken}. Dugan and
Grinstein proposed a formal proof for the 
factorization approach by constructing
a large energy, effective theory \cite {dugan}. 
They established that when the energy of the
generated meson is very large the meson can be factorized out and the deviation
from the factorization amplitude is suppressed by the energy of the factorized
meson.

Furthermore, we will estimate the branching ratios  for the decay modes
$\Lamb\rightarrow n 
(\Lam)\rho^{0}$. In the factorization approach the decay rates for
these processes are determined by the weak matrix elements between $\Lamb$
and $n(\Lam)$. With the aid of heavy quark effective theory (HQET) 
\cite{wise} it is shown that in the heavy quark limit, $m_b \rightarrow
\infty$, there are two independent form factors. We will apply the model
of Refs.\cite{guo1, guo2} to determine these two form factors and 
hence predict the branching ratios for $\Lamb\rightarrow n (\Lam)\rho^{0}$. 

The remainder of this paper is organized as follows. In Sect. II we
give the formalism for the CP-violating asymmetry in 
$\Lamb\rightarrow f \rho^0 (\omega) \rightarrow f \pi^+\pi^-$ ($f=n$ or $\Lam$)
and calculate the strong phases in the factorization approach. Numerical 
results will also be shown in this section. 
In Sect. III we apply the result of HQET and the model
of Refs.\cite{guo1, guo2} to estimate the branching ratios for 
$\Lamb\rightarrow n (\Lam)\rho^{0}$. The results from the nonrelativistic
quark model \cite{cheng1} will also be presented for comparison. 
Finally, Sect. VI is reserved for a brief summary and discussion.

\vspace{0.2in}
{\large\bf II. CP violation in $\Lamb\rightarrow n (\Lam)\pi^+\pi^-$ decays}
\vspace{0.2in}

{\bf II.1 Formalism for CP violation in 
$\Lamb\rightarrow n (\Lam)\pi^+\pi^-$}
\vspace{0.2in}

The formalism for CP violation in $B$ meson hadronic decays 
\cite{eno, tony} can be generalized to $\Lamb$ in a straightforward manner. 
The amplitude, $A$, for the decay $\Lamb \rightarrow f \pi^+ \pi^-$ is:
\begin{equation}
A = \langle \pi^+\pi^- f | {\cal H}^{\rm T} | \Lamb \rangle
+ \langle \pi^+\pi^- f | {\cal H}^{\rm P} | \Lamb \rangle,
\label{2a}
\vspace{2mm}
\end{equation}
where ${\cal H}^{\rm T}$ and ${\cal H}^{\rm P}$ are the 
Hamiltonians for the tree
and penguin diagrams, respectively. Following Refs.\cite{eno, tony}
we define the relative magnitude and phases between these two diagrams 
as follows:
\begin{eqnarray}
A &=& \langle \pi^+\pi^- f | {\cal H}^{\rm T} | \Lamb \rangle \left[
1 + re^{i\delta} e^{i\phi} \right], \nn\\
\bar{A} &=& \langle \pi^+\pi^- \bar{f} | {\cal H}^{\rm T} | \bar{\Lambda}_b
\rangle \left[1 + re^{i\delta} e^{-i\phi} \right],
\label{2b}
\vspace{2mm}
\end{eqnarray}
where $\delta$ and $\phi$ are strong and weak phases, respectively.
$\phi$ is caused by the phase in the CKM matrix, and if the top quark
dominates  penguin diagram contributions it is
arg$[V_{tb}V^{*}_{td}/(V_{ub}V^{*}_{ud})]$ for $b\rightarrow d$
and arg$[V_{tb}V^{*}_{ts}/(V_{ub}V^{*}_{us})]$ for $b\rightarrow s$.
The parameter $r$
is the absolute value of the ratio of tree and penguin amplitudes,
\begin{equation}
r \equiv \left| \frac{\langle \pi^+\pi^- f | {\cal H}^{\rm P} | 
\Lamb \rangle}
{\langle \pi^+\pi^- f | {\cal H}^{\rm T} | \Lamb \rangle}\right|.
\label{2c}
\vspace{2mm}
\end{equation}

The CP-violating asymmetry, $a$, can be written as: 
\begin{equation}
a \equiv { | A |^2 - |{\overline A}|^2 
\over | A |^2 + |{\overline A}|^2 }
= {-2r \sin \delta \sin \phi 
\over 1 + 2r\cos \delta \cos \phi + r^2 }.
\label{2d}
\vspace{2mm}
\end{equation}
It can be seen explicitly from Eq.(\ref{2d}) that both weak and strong
phases are needed to produce CP violation.
$\rho-\omega$ mixing has the dual advantages that the strong phase difference
is large (passing through $90^\circ$ at the $\omega$ resonance) and well known.
In this scenario one has \cite{tony}
\begin{eqnarray}
\langle \pi^+\pi^- f | {\cal H}^{\rm T} | \Lamb \rangle 
&=& {g_{\rho} \over s_\rho s_\omega} \tilde\Pi_{\rho\omega} t_{\omega}
  + { g_{\rho} \over s_\rho } t_\rho, \label{2e1}\\
\langle \pi^+\pi^- f | {\cal H}^{\rm P} | \Lamb \rangle 
&=& {g_{\rho} \over s_\rho s_\omega} \tilde\Pi_{\rho\omega} p_{\omega}
  + { g_{\rho} \over s_\rho } p_\rho, 
\label{2e2}
\vspace{2mm}
\end{eqnarray}
where $t_{\rm V}$ (V=$\rho$ or $\omega$) is the tree 
and $p_{\rm V}$ is the penguin amplitude for 
producing a vector meson, ${\rm V}$, by $\Lamb \rightarrow f {\rm V}$,
$g_\rho$ is the coupling for $\rho^0 \ra \pi^+\pi^-$, 
$\tilde\Pi_{\rho\omega}$ is the effective  $\rho - \omega$ mixing amplitude,
and $s_{\rm V}^{-1}$ is the propagator of V,
\begin{equation}
s_{\rm V}=s - m_{\rm V}^2 + i m_{\rm V} \Gamma_{\rm V},
\label{2f}
\vspace{2mm}
\end{equation}
with $\sqrt{s}$ being the invariant mass of the $\pi^+ \pi^-$ pair.

$\tilde\Pi_{\rho\omega}$ is extracted \cite{pionff97} from 
the data for $e^+e^-\ra \pi^+\pi^-$ \cite{barkov85} when 
$\sqrt s$ is near the $\omega$ mass. Detailed discussions can be found in
Refs.\cite{tony, pionff97, pionffn97}. The numerical values are 
$${\rm Re}\tilde\Pi_{\rho\omega}(m_\omega^2) = -3500\pm 300
{\rm MeV}^2,\;\; {\rm Im}\tilde\Pi_{\rho\omega}(m_\omega^2)= -300 
\pm 300 {\rm MeV}^2.$$
We stress that the direct coupling $\omega \ra \pi^+\pi^-$ is effectively
absorbed into $\tilde\Pi_{\rho\omega}$, where it contributes some 
$s$-dependence. The limits on this $s$-dependence, 
$\tilde\Pi_{\rho\omega}(s)=
\tilde\Pi_{\rho\omega}(m_\omega^2)
+ (s - m_\omega^2)\tilde\Pi_{\rho\omega}'(m_\omega^2)$,
were determined in the fit of Gardner and O'Connell,
$\tilde\Pi_{\rho\omega}'(m_\omega^2) = 0.03 \pm 0.04$~\cite{pionff97}.
In practice, the effect of the derivative term is negligible.

From Eqs.(\ref{2a},\ref{2b},\ref{2e1},\ref{2e2}) one has
\begin{equation}
re^{i\delta}\,e^{i\phi}= { \tilde\Pi_{\rho\omega} p_{\omega} + s_\omega p_\rho
\over
\tilde\Pi_{\rho\omega} t_{\omega} + s_\omega t_\rho}.
\label{2g}
\vspace{2mm}
\end{equation}
Defining
\begin{equation}
{p_\omega \over t_\rho} \equiv r' e^{i(\delta_q + \phi)}, \quad
{t_\omega \over t_\rho} \equiv \alpha e^{i \delta_\alpha}, \quad
{p_\rho \over p_\omega} \equiv \beta e^{i \delta_\beta},
\label{2h}
\vspace{2mm}
\end{equation}
where $\delta_\alpha$, $\delta_\beta$ and $\delta_q$ are strong phases, one
has the following expression from Eq.(\ref{2g})
\begin{equation}
re^{i\delta} = r' e^{i\delta_q} \frac{
 \tilde\Pi_{\rho\omega} + \beta e^{i\delta_\beta}s_\omega}{s_\omega
+\tilde\Pi_{\rho\omega} \alpha e^{i\delta_\alpha}}.
\label{2i}
\vspace{2mm}
\end{equation}

It will be shown that in the factorization approach, for both 
$\Lamb\rightarrow n \pi^+\pi^-$ and $\Lamb\rightarrow \Lam\pi^+\pi^-$,
we have (see II.3 for details)
\begin{equation}
\alpha e^{i\delta_\alpha}=1.
\label{2j}
\vspace{2mm}
\end{equation}
Letting
\begin{equation}
\beta e^{i\delta_\beta}=b+ci,\;\;r' e^{i\delta_q}=d+ei, 
\label{2k}
\vspace{2mm}
\end{equation}
and using Eq.(\ref{2i}), we obtain the following result when $\sqrt{s}\sim
m_\omega$,
\begin{equation}
re^{i\delta} = \frac{C+Di}{(s-m_{\omega}^{2}+{\rm Re}\tilde\Pi_{\rho\omega})^2
+({\rm Im}\tilde\Pi_{\rho\omega}+m_\omega \Gamma_\omega)^2},
\label{2l}
\vspace{2mm}
\end{equation}
where
\begin{eqnarray}
C&=&(s-m_{\omega}^{2}+{\rm Re}\tilde\Pi_{\rho\omega})\{d
[{\rm Re}\tilde\Pi_{\rho\omega}+b(s-m_{\omega}^{2})-cm_\omega \Gamma_\omega]
\nn\\
&&
-e[{\rm Im}\tilde\Pi_{\rho\omega}+bm_\omega \Gamma_\omega+c(s-m_{\omega}^{2})]
\} \nn\\
& &+({\rm Im}\tilde\Pi_{\rho\omega}+m_\omega \Gamma_\omega)
\{e[{\rm Re}\tilde\Pi_{\rho\omega}+b(s-m_{\omega}^{2})-cm_\omega \Gamma_\omega]
\nn\\
&&+d[{\rm Im}\tilde\Pi_{\rho\omega}+bm_\omega \Gamma_\omega+c(s-m_{\omega}^{2})]
\}, \nn\\
D&=&(s-m_{\omega}^{2}+{\rm Re}\tilde\Pi_{\rho\omega})\{e
[{\rm Re}\tilde\Pi_{\rho\omega}+b(s-m_{\omega}^{2})-cm_\omega \Gamma_\omega]
\nn\\
&&+d[{\rm Im}\tilde\Pi_{\rho\omega}+bm_\omega \Gamma_\omega+c(s-m_{\omega}^{2})]
\} \nn\\
& &-({\rm Im}\tilde\Pi_{\rho\omega}+m_\omega \Gamma_\omega)
\{d[{\rm Re}\tilde\Pi_{\rho\omega}+b(s-m_{\omega}^{2})-cm_\omega \Gamma_\omega]
\nn\\
&&
-e[{\rm Im}\tilde\Pi_{\rho\omega}+bm_\omega \Gamma_\omega+c(s-m_{\omega}^{2})]
\}. 
\label{2m}
\vspace{2mm}
\end{eqnarray}

$\beta e^{i\delta_\beta}$ and $r' e^{i\delta_q}$ will be calculated later. Then
from Eqs.(\ref{2l}) and (\ref{2m}) 
we obtain $r{\rm sin}\delta$, $r{\rm cos}\delta$ and $r$. 
In order to get the CP-violating asymmetry $a$ in Eq.(\ref{2d}) 
${\rm sin}\phi$ and  ${\rm cos}\phi$ are needed. $\phi$ is determined
by the CKM matrix elements. In the Wolfenstein parametrization \cite{wolf},
and in the approximation that the top quark dominates the penguin diagrams,
we have 
\begin{eqnarray}
({\rm sin}\phi)^n&=&\frac{\eta}{\sqrt{[\rho (1-\rho)-\eta^2]^2+\eta^2}}, \nn\\
({\rm cos}\phi)^n&=&\frac{\rho (1-\rho)-\eta^2}{\sqrt{[\rho (1-\rho)-\eta^2]^2
+\eta^2}}, 
\label{2n1}
\vspace{2mm}
\end{eqnarray}
for $\Lamb\rightarrow n \pi^+\pi^-$, and
\begin{eqnarray}
({\rm sin}\phi)^\Lam
&=&-\frac{\eta}{\sqrt{[\rho (1+\lambda^2\rho)+\lambda^2\eta^2]^2
+\eta^2}}, \nn\\
({\rm cos}\phi)^\Lam&=&-\frac{\rho (1+\lambda^2\rho)+\lambda^2\eta^2} 
{\sqrt{[\rho (1+\lambda^2\rho)+\lambda^2\eta^2]^2
+\eta^2}},
\label{2n2}
\vspace{2mm}
\end{eqnarray}
for $\Lamb\rightarrow \Lam \pi^+\pi^-$. Note that here, and in what follows, 
all the quantities
with the superscript $n$ (or $\Lam$) represent those for
$\Lamb\rightarrow n \rho^0 (\omega)$ (or
$\Lamb\rightarrow \Lam \rho^0 (\omega))$.

\vspace{0.2in}
{\bf II.2 The effective Hamiltonian} 
\vspace{0.2in}

With the operator product expansion, the 
effective Hamiltonian relevant to the
processes $\Lamb\rightarrow f \rho^0 (\omega)$ is
\begin{equation}
H_{\Delta B=1} = {G_F\over \sqrt{2}}[V_{ub}V^*_{uq}(c_1O^u_1 + c_2 O^u_2)
 - V_{tb}V^*_{tq}\sum^{10}_{i=3} c_iO_i] +H.C.\;,
\label{2o}
\vspace{2mm}
\end{equation}
where the Wilson coefficients, $c_i\;(i=1,...,10)$, are calculable
in perturbation theory and are scale dependent. They are defined at 
the scale $\mu\approx m_b$ in our case. The quark $q$ could be $d$ or
$s$ for our purpose. The operators $O_i$ have the following expression   
\begin{eqnarray}
O^u_1&=& \bar q_\alpha \gamma_\mu(1-\gamma_5)u_\beta\bar
u_\beta\gamma^\mu(1-\gamma_5)b_\alpha,\;\;
O^u_2= \bar q \gamma_\mu(1-\gamma_5)u\bar
u\gamma^\mu(1-\gamma_5)b,\nn\\
O_3&=& \bar q \gamma_\mu(1-\gamma_5)b \sum_{q'}
\bar q' \gamma^\mu(1-\gamma_5) q',\;\;
O_4 = \bar q_\alpha \gamma_\mu(1-\gamma_5)b_\beta \sum_{q'}
\bar q'_\beta \gamma^\mu(1-\gamma_5) q'_\alpha,\nn\\
O_5&=&\bar q \gamma_\mu(1-\gamma_5)b \sum_{q'} \bar q'
\gamma^\mu(1+\gamma_5)q',\;\;
O_6 = \bar q_\alpha \gamma_\mu(1-\gamma_5)b_\beta \sum_{q'}
\bar q'_\beta \gamma^\mu(1+\gamma_5) q'_\alpha,\nn\\
O_7&=&{3\over 2}\bar q \gamma_\mu(1-\gamma_5)b \sum_{q'} e_{q'}\bar q'
\gamma^\mu(1+\gamma_5)q',\nn\\
O_8&=& {3\over 2}\bar q_\alpha \gamma_\mu(1-\gamma_5)b_\beta \sum_{q'}
e_{q'}\bar q'_\beta \gamma^\mu(1+\gamma_5) q'_\alpha,\nn\\
O_9&=&{3\over 2}\bar q \gamma_\mu(1-\gamma_5)b \sum_{q'} e_{q'}\bar q'
\gamma^\mu(1-\gamma_5)q',\nn\\
O_{10}&=& {3\over 2}\bar q_\alpha \gamma_\mu(1-\gamma_5)b_\beta \sum_{q'}
e_{q'}\bar q'_\beta \gamma^\mu(1-\gamma_5) q'_\alpha,
\label{2p}
\vspace{2mm}
\end{eqnarray}
where $\alpha$ and $\beta$ are color indices, and 
$q'=u,\;d,\;s$ quarks. In Eq.(\ref{2p})
$O_1$ and $O_2$ are the tree level operators. $O_{3}-O_{6}$ are QCD 
penguin operators, which are isosinglet. 
$O_{7}-O_{10}$ arise from electroweak penguin diagrams, and they have 
both isospin 0 and 1 components.

The Wilson coefficients, $c_i$, are known to the next-to-leading logarithmic
order \cite{he, buras}. At the scale $\mu=m_b=5{\rm GeV}$ their values are
\begin{eqnarray}
c_1 &=& -0.3125, \;\;c_2 = 1.1502, \;\;c_3 = 0.0174, \;\;c_4 = -0.0373,
\nn\\
c_5 &=& 0.0104,\;\;c_6 = -0.0459,\;\; c_7 = -1.050\times 10^{-5},\nn\\
c_8 &=& 3.839\times 10^{-4},
\;\;c_9 = -0.0101, \;\;c_{10} = 1.959\times 10^{-3}.
\label{2p1}
\vspace{2mm}
\end{eqnarray}
To be consistent, the matrix elements of the operators
$O_i$ should also be renormalized to the one-loop order. This results in 
the effective Wilson coefficients, $c'_i$, which satisfy the constraint
\begin{equation}
c_i (\mu) \langle O_i (\mu) \rangle =c'_i \langle O_{i}^{\rm tree} \rangle,
\label{2q}
\vspace{2mm}
\end{equation}
where $\langle O_i (\mu) \rangle$ are the matrix elements, renormalized to 
one-loop order.  The relations between $c'_i$ and $c_i$ read
\begin{eqnarray}
c'_1 &=& c_1,\;\; c'_2 = c_2, \;\;c'_3 =  c_3 - P_s/3, \nn\\
c'_4 &=&  c_4 +P_s,\;\;c'_5 = c_5 - P_s/3,\;\;c'_6 = c_6 + P_s,\nn\\
c'_7 &=&  c_7 +P_e,\;\;c'_8 =c_8,\;\;
c'_9 = c_9 +P_e,\;\;c'_{10} =  c_{10},
\label{2r}
\vspace{2mm}
\end{eqnarray}
where 
$$P_s = (\alpha_s/8\pi) c_2 (10/9 +G(m_c,\mu,q^2)),\;\;
P_e = (\alpha_{em}/9\pi)(3 c_1+ c_2) (10/9 + G(m_c,\mu,q^2)),$$
with 
$$G(m_c,\mu,q^2) = 4\int^1_0 {\rm d}x x(1-x) {\rm ln}{m_{c}^{2}-x(1-x)q^2\over
\mu^2},$$
where $q^2$ is the momentum transfer of the gluon or photon in the penguin
diagrams. $G(m_c,\mu,q^2)$ has the following explicit expression \cite{kramer}
\begin{eqnarray}
{\rm Re}G&=&\frac{2}{3}\left({\rm ln}\frac{m_{c}^2}{\mu^2}-\frac{5}{3}
-4\frac{m_{c}^2}{q^2}+(1+2\frac{m_{c}^2}{q^2})\sqrt{1-4\frac{m_{c}^2}{q^2}}
{\rm ln}\frac{1+\sqrt{1-4\frac{m_{c}^2}{q^2}}}{1-\sqrt{1-4\frac{m_{c}^2}{q^2}}}
\right),\nn\\
{\rm Im}G&=&-\frac{2}{3}\pi\left(1+2\frac{m_{c}^2}{q^2}\right)
\sqrt{1-4\frac{m_{c}^2}{q^2}}.
\label{2s}
\vspace{2mm}
\end{eqnarray}

Based on simple arguments for $q^2$ at the quark level, the value
of $q^2$ is chosen in the range $0.3 < q^2/m_{b}^2 < 0.5$\cite{eno, tony}. 
From Eqs.(\ref{2p1}), (\ref{2r}) and (\ref{2s}) we can obtain 
numerical values of $c'_i$. When $q^2/m_{b}^2=0.3$,
\begin{eqnarray}
c'_1 &=&-0.3125, \;\; c'_2=1.1502\nn\\
c'_3 &=& 2.433\times 10^{-2}+1.543\times 10^{-3}i,\;\; 
c'_4 = -5.808\times 10^{-2}-4.628\times 10^{-3}i,\nn\\ 
c'_5 &=& 1.733\times 10^{-2}+1.543\times 10^{-3}i,\;\; 
c'_6 = -6.668\times 10^{-2}-4.628\times 10^{-3}i,\nn\\
c'_7 &=& -1.435\times 10^{-4}-2.963\times 10^{-5}i,\;\;
c'_8 = 3.839 \times 10^{-4},\nn\\
c'_9 &=& -1.023\times 10^{-2}-2.963\times 10^{-5}i,\;\;
c'_{10} = 1.959 \times 10^{-3},
\label{2c1}
\vspace{2mm}
\end{eqnarray}
and when $q^2/m_{b}^2=0.5$,
\begin{eqnarray}
c'_1 &=&-0.3125, \;\; c'_2=1.1502\nn\\
c'_3 &=& 2.120\times 10^{-2}+5.174\times 10^{-3}i,\;\; 
c'_4 = -4.869\times 10^{-2}-1.552\times 10^{-2}i,\nn\\ 
c'_5 &=& 1.420\times 10^{-2}+5.174\times 10^{-3}i,\;\; 
c'_6 = -5.729\times 10^{-2}-1.552\times 10^{-2}i,\nn\\
c'_7 &=& -8.340\times 10^{-5}-9.938\times 10^{-5}i,\;\;
c'_8 = 3.839 \times 10^{-4},\nn\\
c'_9 &=& -1.017\times 10^{-2}-9.938\times 10^{-5}i,\;\;
c'_{10} = 1.959 \times 10^{-3},
\label{2c2}
\vspace{2mm}
\end{eqnarray}
where we have taken $\alpha_s (m_Z) =0.112$, $\alpha_{em} (m_b)=1/132.2$,
$m_b=5$GeV and $m_c=1.35$GeV.

\vspace{0.2in}
{\bf II.3 CP violation in $\Lamb\rightarrow n (\Lam) \pi^+\pi^-$} 
\vspace{0.2in}

In the following we will calculate the CP-violating asymmetries 
in $\Lamb\rightarrow n (\Lam)\pi^+\pi^-$. 
With the Hamiltonian in Eq.(\ref{2o}) we are ready to
evaluate the matrix elements. In the factorization approximation
$\rho^0 (\omega)$ is generated by one current which has the proper
quantum numbers in the Hamiltonian. 

First we consider $\Lamb\rightarrow n \rho^0 (\omega)$. After factorization,
the contribution to $t^{n}_{\rho}$ from the tree level operator $O_1$ is
\begin{equation}
\langle \rho^0 n |O_1 |\Lamb \rangle = \langle \rho^0 |(\bar{u}u)|0\rangle
\langle n | (\bar{d} b) |\Lamb \rangle \equiv T,
\label{2t}
\vspace{2mm}
\end{equation}
where $(\bar{u}u)$ and $(\bar{d} b)$ denote the V-A currents. 
Using the Fierz transformation the contribution of $O_2$ is $\frac{1}{N_c} T$.
Hence we have
\begin{equation}
t^{n}_{\rho}=(c_1+\frac{1}{N_c} c_2)T.
\label{2u}
\vspace{2mm}
\end{equation}
It should be noted that in Eq.(\ref{2u}) we have neglected the color-octet 
contribution, which is nonfactorizable and difficult to calculate.
Therefore, $N_c$ should be treated as an effective parameter and may
deviate from the naive value 3. 
In the same way we find that $t^{n}_{\omega} = t^{n}_{\rho}$, 
hence from Eq.(\ref{2h})
\begin{equation}
(\alpha e^{i\delta_\alpha})^n=1.
\label{2v1}
\vspace{2mm}
\end{equation}

In a similar way, we can evaluate the penguin operator contributions
$p^{n}_{\rho}$ and $p^{n}_{\omega}$ with the aid of the Fierz identities.
From  Eq.(\ref{2h}) we have
\begin{equation}
(\beta e^{i\delta_\beta})^n=
\frac{-2(c'_4+\frac{1}{N_c}c'_3)
+3(c'_7+\frac{1}{N_c}c'_8)+(3+\frac{1}{N_c})c'_9+(1+\frac{3}{N_c})c'_{10}}
{2(2+\frac{1}{N_c})c'_3+2(1+\frac{2}{N_c})c'_4+4(c'_5+\frac{1}{N_c}c'_6)
+c'_7+\frac{1}{N_c}c'_8+(c'_9-c'_{10})(1-\frac{1}{N_c})}, \nn\\
\label{2v2}\\
\vspace{2mm}
\end{equation}
\begin{eqnarray}
(r' e^{i\delta_q})^n&=&-\frac{2(2+\frac{1}{N_c})c'_3+2(1+\frac{2}{N_c})c'_4
+4(c'_5+\frac{1}{N_c}c'_6)+c'_7+\frac{1}{N_c}c'_8+
(c'_9-c'_{10})(1-\frac{1}{N_c})}{
2(c_1+\frac{1}{N_c}c_2)}\nn\\
&&\cdot\left|\frac{V_{tb}V^{*}_{td}}{V_{ub}V^{*}_{ud}}\right|,
\label{2v3}
\vspace{2mm}
\end{eqnarray}
where 
\begin{equation}
\left|\frac{V_{tb}V^{*}_{td}}{V_{ub}V^{*}_{ud}}\right|=\frac{\sqrt{[\rho
(1-\rho)-\eta^2]^2+\eta^2}}{(1-\lambda^2/2)(\rho^2+\eta^2)}.
\label{2x}
\vspace{2mm}
\end{equation}

For $\Lamb\rightarrow \Lam \rho^0 (\omega)$, the evaluation is same. 
Defining
\begin{equation}
\langle \rho^0 \Lam |O_1 |\Lamb \rangle = \langle \rho^0 |(\bar{u}u)|0\rangle
\langle \Lam | (\bar{s} b) |\Lamb \rangle \equiv \tilde{T},
\label{2t2}
\vspace{2mm}
\end{equation}
we have
\begin{equation}
t^{\Lam}_{\rho}=(c_1+\frac{1}{N_c} c_2)\tilde{T}.
\label{2u2}
\vspace{2mm}
\end{equation}

After evaluating the penguin diagram contributions we obtain the
following results,
\begin{equation}
(\alpha e^{i\delta_\alpha})^\Lam=1,
\label{2w1}
\vspace{2mm}
\end{equation}
\begin{equation}
(\beta e^{i\delta_\beta})^\Lam=
\frac{3(c'_7+\frac{1}{N_c}c'_8+c'_9+\frac{1}{N_c}c'_{10})}
{4(c'_3+\frac{1}{N_c}c'_4+c'_5+\frac{1}{N_c}c'_6)+
c'_7+\frac{1}{N_c}c'_8+c'_9+\frac{1}{N_c}c'_{10}},
\label{2w2}\\
\vspace{2mm}
\end{equation}
\begin{equation}
(r' e^{i\delta_q})^\Lam=-\frac{4(c'_3+\frac{1}{N_c}c'_4+c'_5
+\frac{1}{N_c}c'_6)+c'_7+\frac{1}{N_c}c'_8+c'_9+\frac{1}{N_c}c'_{10}}
{2(c_1+\frac{1}{N_c}c_2)}
\left|\frac{V_{tb}V^{*}_{ts}}{V_{ub}V^{*}_{us}}\right|,
\label{2w3}
\vspace{2mm}
\end{equation}
where 
\begin{equation}
\left|\frac{V_{tb}V^{*}_{ts}}{V_{ub}V^{*}_{us}}\right|=\frac{\sqrt{[\rho
(1+\lambda^2\rho)+\lambda^2\eta^2]^2+\eta^2}}{\lambda^2(\rho^2+\eta^2)}.
\label{2y}
\vspace{2mm}
\end{equation}

It can be seen from Eqs.(\ref{2v2}) and 
(\ref{2w2}) that $\beta$ and $\delta_\beta$
are determined solely by the Wilson coefficients. On the other hand,
$r'$ and $\delta_q$ depend on both the Wilson coefficients and the CKM 
matrix elements, as shown in Eqs.(\ref{2v3}) and (\ref{2w3}).
Substituting Eqs.(\ref{2v1}, \ref{2v2}, \ref{2v3}, \ref{2w1}, \ref{2w2}, 
\ref{2w3}) into Eqs.(\ref{2k}, \ref{2l}, \ref{2m}) we can obtain 
$(r {\rm sin}\delta)^{n(\Lam)}$ and $(r {\rm cos}\delta)^{n(\Lam)}$.
Then in combination with with Eqs.(\ref{2n1}) and (\ref{2n2}) the CP-violating
asymmetries $a$ can be obtained.

In the numerical calculations, we have several parameters: $q^2$, $N_c$,
and the CKM matrix elements in the Wolfenstein parametrization.
As mentioned in II.2, the value
of $q^2$ is chosen in the range $0.3 < q^2/m_{b}^2 < 0.5$\cite{eno, tony}. 

The CKM matrix elements
should be determined from experiment. $\lambda$ is well measured
\cite{leu} and we will use $\lambda=0.221$ in our numerical calculations. 
However, due to the large experimental errors at present, 
$\rho$ and $\eta$ are not yet fixed. 
From $b \ra u$ transitions $\sqrt{\rho^2+\lambda^2}=0.363\pm0.073$
\cite{neu, he2}. In combination with the results from $B^0-\bar{B}^0$
mixing\cite{rosner} we have $0.18 < \eta < 0.42$\cite{he2}. In our calculations
we use $\eta=0.34$ as in Refs.\cite{eno, tony}. Recently, 
it has been pointed out\cite{cheng2}
that from the branching ratio of $B^\pm \ra \eta \pi^\pm$ a negative value
for $\rho$ is favored. Hence we will use $\rho=-0.12$, corresponding to
$\eta=0.34$.  These values lead to $\phi^n=126^\circ$ and 
$\phi^\Lambda=-72^\circ$ from Eqs.(\ref{2n1}) and (\ref{2n2}).

The value of the effective $N_c$ should also be determined by experiments.
The analysis of the data for $B \ra D \pi$,
$B^\pm \ra \omega \pi^\pm$ and
$B^\pm \ra \omega K^\pm$ indicates that $N_c$ is about 2 
\cite{cheng3, neubert2}. For the $\Lamb$
decays, we do not have enough data to extract $N_c$ at present. 
Finally, we use $m_b=5$GeV, $m_c=1.35$GeV, $\alpha_s (m_Z)
=0.112$ and $\alpha_{em} (m_b)=1/132.2$ to calculate the Wilson coefficients,
$c'_i$, as discussed in II.2 (see Eqs.(\ref{2c1}) and (\ref{2c2})).
The numerical values of $\beta,\;r',\;\delta_\beta$ 
and $\delta_q$ for $\Lamb \ra n \rho^0$ and $\Lamb \ra \Lam \rho^0$
are listed in Tables 1 and 2, respectively.

\begin{table}
\caption{Values  of  $\beta,\;r',\;\delta_\beta$ 
and $\delta_q$ for $\Lamb \ra n \rho^0$}
\begin{center}
\begin{tabular}{lccccc}
\hline
\hline
$N_c$&$q^2/m_{b}^{2}$&$\beta$&$r'$ &$\delta_\beta$ &$\delta_q$\\
\hline
\hline
2&0.3&0.339&1.149&-3.096&0.0769\\
\hline
2&0.5&0.328&1.011&-2.935&0.297\\
\hline
3&0.3&0.649&2.537&-3.103&0.0766\\
\hline
3&0.5&0.629&2.233&-2.970&0.296\\
\hline
\hline
\end{tabular}
\end{center}
\end{table}

\begin{table}
\caption{Values  of  $\beta,\;r',\;\delta_\beta$ 
and $\delta_q$ for $\Lamb \ra \Lam \rho^0$}
\begin{center}
\begin{tabular}{lccccc}
\hline
\hline
$N_c$&$q^2/m_{b}^{2}$&$\beta$&$r'$ &$\delta_\beta$ &$\delta_q$\\
\hline
\hline
2&0.3&0.299&9.925&-0.0611&0.0675\\
\hline
2&0.5&0.332&8.833&-0.235&0.257\\
\hline
3&0.3&3.086&3.715&-1.766$\times 10^{-4}$&6.353$\times 10^{-3}$\\
\hline
3&0.5&3.087&3.668&-6.071$\times 10^{-4}$&0.0216\\
\hline
\hline
\end{tabular}
\end{center}
\end{table}

In Figs.1 and 2 we plot the numerical values of the CP-violating 
asymmetries,
$a$, for $\Lamb\rightarrow n \pi^+\pi^-$ and $\Lamb\rightarrow \Lam\pi^+\pi^-$,
respectively, for $N_c=2$. It can be seen that there is a very
large CP violation 
when the invariant mass of the $\pi^+\pi^-$ is near the $\omega$ mass. 
For $\Lamb\rightarrow n \pi^+\pi^-$ the maximum CP-violating asymmetry
is $a_{max}^{n}=-66\%$ ($q^2/m_{b}^{2}=0.3$) and 
$a_{max}^{n}=-50\%$ ($q^2/m_{b}^{2}=0.5$), while
for  $\Lamb\rightarrow \Lam\pi^+\pi^-$, $a_{max}^{\Lam}=
68\%$ ($q^2/m_{b}^{2}=0.3$) and $a_{max}^{\Lam}=76\%$ ($q^2/m_{b}^{2}=0.5$).
It would be very interesting to actually 
measure such large CP-violating asymmetries.

Although $N_c$ is around 2 for $B$ decays, it might be different in the
$\Lamb$ case. We also calculated the numerical values when $N_c=3$.
It is found that, in this case, we still have large CP violation for
$\Lamb\rightarrow n \pi^+\pi^-$, with 
$a_{max}^{n}=-52\%$ ($q^2/m_{b}^{2}=0.3$) and 
$a_{max}^{n}=-40\%$ ($q^2/m_{b}^{2}=0.5$). However, for
$\Lamb\rightarrow \Lam\pi^+\pi^-$, $a_{max}^{\Lam}$ is much smaller, 
only about $6\%$.


\vspace{0.2in}
{\large\bf III. Branching ratios for $\Lamb\rightarrow n (\Lam)\rho^0$}
\vspace{0.2in}

In this section we estimate the branching ratios for 
$\Lamb\rightarrow f \rho^0$. In the factorization
approach, $\rho^0$ is factorized out and hence the decay amplitude is 
determined by the  weak transition matrix elements $\Lamb \ra f$.
In the heavy quark limit, $m_b \ra \infty$,  
it is shown in the HQET that there are two form factors for 
$\Lamb \ra f$ \cite{roberts},
\begin{equation}
\langle f (p_f) | \bar{q}\gamma_\mu (1-\gamma_5) b | \Lamb (v) \rangle
=\bar{u}_f (p_f)[F_1(v\cdot p_f)+\rlap/v F_2(v\cdot p_f)]\gamma_\mu 
(1-\gamma_5) u_{\Lamb} (v),
\label{3a}
\vspace{2mm}
\end{equation}
where $q=d$ or $s$;
$u_f$ and $u_{\Lamb}$ are the Dirac spinors of $f$ and $\Lamb$, 
respectively; $p_f$ is the momentum of the final baryon, $f$, and $v$ is the 
velocity of $\Lamb$. In order to calculate $F_1$ and $F_2$ we need two 
constraints.

In Ref.\cite{stech} the author proposed two dynamical assumptions with respect
to the meson structure and decays: (i) in the rest frame of a hadron the 
distribution of the off-shell momentum components of the constituents
is strongly peaked at 
zero with a width of the order of the 
confinement scale; (ii) during the weak transition
the spectator retains its momentum and spin. These two assumptions led to the 
result that the matrix element of the heavy to light meson transition
is dominated by the configuration where the active quarks' momenta are almost
equal to those of their corresponding mesons. This argument is corrected by
terms of order $1/m_b$ and $\Lambda_{\rm QCD}/E_f$, 
and hence is a good approximation
in heavy hadron decays.   
Some relations among the form factors in the heavy to light meson transitions
are found in this approximation. In Ref.\cite{guo1} the above approach is
generalized to the baryon case and a relation between $F_1$ and $F_2$ is 
found.

Another relation between $F_1$ and $F_2$ comes from the overlap integral of the
hadronic wave functions of $\Lamb$ and $f$. In the heavy quark limit $\Lamb$
is regarded as a bound state of a heavy quark $b$ and a light scalar diquark
$[ud]$\cite{guo1, guo2, guo3}. On the other hand,
the light baryon $f$ has various quark-diquark configurations\cite{kroll}
and only the $q[ud]$ component contributes to the transition $\Lamb \ra f$.
This leads to a suppression 
factor, $C_s$, which is the Clebsch-Gordan coefficient
of $q[ud]$. $C_s=1/\sqrt{2}$ for $n$ and $C_s=1/\sqrt{3}$ for $\Lam$,
respectively\cite{kroll}. 
In the quark-diquark picture, the hadronic wave function has
the following form
\begin{equation}
\psi_{i}(x_{1},\vec{k}_{\perp}) = N_{i}x_{1}x_{2}^3 {\rm exp} [-b^{2}
(\vec{k}_{\perp}^2+ m_{i}^{2}(x_{1}-x_{0i})^{2})],
\label{3b}
\vspace{2mm}
\end{equation}
where $i=\Lamb$, $n$ or $\Lam$; $x_1,\; x_2$ ($x_2=1-x_1)$ 
are the longitudinal 
momentum fractions of the active quark and the diquark, respectively;
$\vec{k}_{\perp}$ is the transverse momentum; $N_i$ is the normalization
constant; the parameter $b$ is related to the average transverse momentum,
$b$=1.77GeV and $b$=1.18GeV, corresponding to 
$\langle k_{\perp}^{2}\rangle$$^{\frac{1}{2}}$ = 400 MeV and 
$\langle k_{\perp}^{2}\rangle$$^{\frac{1}{2}}$ = 600MeV respectively;
and $x_{0i}$ ($x_{0i}=1-\epsilon/m_i$, $\epsilon$ is the 
mass of the diquark) is the
peak position of the wave function.
By working in the appropriate infinite momentum frame
and evaluating the good current components\cite{guo1, guo2},
another relation between $F_1$ and $F_2$ is given in terms of the overlap 
integral of the hadronic wave functions of $\Lamb$ and $f$. Therefore,
$F_1$ and $F_2$ are obtained as the following,
\begin{eqnarray}
F_{1}&=& \frac{2E_{f}+m_{f}+m_{q}}{2(E_{f}+m_{q})}C_{s}I(\omega),\nn\\
F_{2}&=& \frac{m_{q}-m_{f}}{2(E_{f}+m_{q})}C_{s}I(\omega),
\label{3c}
\vspace{2mm}
\end{eqnarray}
where $I(\omega)$ is the overlap integral of the hadronic wave functions of $\Lamb$ and $f$,
\begin{eqnarray}
I(\omega) &=& \left(\frac{2}{\omega+1}\right)^{7/4}y^{-9/2}
[A_f K_6(\sqrt{2} b \epsilon)]^{-1/2}
{\rm exp}\left(-2b^2\epsilon^2\frac{\omega-1}
{\omega+1}\right)\nn\\
&&\int_{-\frac{2b\epsilon}{\sqrt{\omega+1}}}^{y
-\frac{2b\epsilon}{\sqrt{\omega+1}}}{\rm d}z 
\;{\rm exp}(-z^2)
\left(y-\frac{2b\epsilon}{\sqrt{\omega+1}}-z\right)
\left(z+\frac{2b\epsilon}{\sqrt{\omega+1}}\right)^6,
\label{3d}
\vspace{2mm}
\end{eqnarray}
and $y=bm_f\sqrt{\omega+1}$, with $\omega$ being the velocity transfer 
$\omega=v\cdot p_f/m_f$ and $A_f$ and $K_6$ defined as 
\begin{eqnarray}
A_f&=&\int_{0}^{1}{\rm d}x\;x^6 (1-x)^2{\rm exp}[-2b^2 m_{f}^{2} 
(x-\epsilon/m_f)^2],\nn\\
K_6(\sqrt{2} b \epsilon)&=&\int_{-\sqrt{2} b \epsilon}^{\infty}
{\rm d }x \;{\exp}(-x^2)(x+\sqrt{2} b \epsilon)^6.
\label{3e}
\vspace{2mm}
\end{eqnarray}
It should be noted that in Eqs.(\ref{3d}) and (\ref{3e}) we have taken the limit 
$m_b \ra \infty$. 

It can be shown that $\omega=3.03$ for $\Lamb \ra n \rho^0$ and $\omega=2.58$
for $\Lamb \ra \Lam \rho^0$. 
Taking $\epsilon=600$MeV, from Eq.(\ref{3d}), we find that 
$I^{n}=0.0258 (0.0509)$ for $b$=1.77GeV$^{-1}$ ($b$=1.18GeV$^{-1}$), and
$I^{\Lam}=0.0389 (0.0781)$ for $b$=1.77GeV$^{-1}$ ($b$=1.18GeV$^{-1}$). 
Substituting these
numbers into Eq.(\ref{3c}) we obtain the following results,
\begin{eqnarray}
F_{1}^{n}&=&-0.0199 \;(-0.0393), \;\;{\rm for}\;\; b=1.77 (1.18)
{\rm GeV}^{-1}, \nn\\
F_{2}^{n}&=&0.00168 \;(0.00332), \;\;{\rm for}\;\; 
b=1.77 (1.18){\rm GeV}^{-1},
\label{3f}
\vspace{2mm}
\end{eqnarray}
and
\begin{eqnarray}
F_{1}^{\Lam}&=&0.0245 \;(0.0492),\;\; {\rm for} \;\;b=1.77 (1.18)
{\rm GeV}^{-1},\nn\\
F_{2}^{\Lam}&=&-0.00205 \;(-0.00411), \;\;{\rm for} \;\;b=1.77 (1.18)
{\rm GeV}^{-1},
\label{3g}
\vspace{2mm}
\end{eqnarray}
where we have taken $m_d=0.35$GeV and $m_s=0.50$GeV.

To estimate the branching ratios for $\Lamb \ra n (\Lam) \rho^0$ we only
take the $O_1$ and $O_2$ terms in the Hamiltonian (\ref{2o}), since they
give dominant contributions. In the factorization approach, the amplitude
for $\Lamb \ra n (\Lam) \rho^0$ has the following form
\begin{equation}
A(\Lambda_{b}\rightarrow f + \rho^0) = \frac{G_{F}}{\sqrt{2}}V_{ub}
  V_{uq}^{*}a_{1}\langle\rho^0\mid\bar{u}\gamma_{\mu}(1-\gamma_{5})u\mid 0
\rangle
  \langle f\mid\bar{q}\gamma^{\mu}(1-\gamma_{5})b\mid\Lambda_{b}\rangle,
\label{3h}
\vspace{2mm}
\end{equation}
where 
\begin{equation}
a_1=c_1+ \frac{1}{N_c} c_2.
\label{3i}
\vspace{2mm}
\end{equation}
In Eq.(\ref{3h}) $\rho^0$ has been factorized out and the matrix element
$\langle\rho^0\mid\bar{u}\gamma_{\mu}(1-\gamma_{5})u\mid 0 \rangle$ is related
to the decay constant $f_\rho$. From Eq.(\ref{3h}) the branching ratios
for $\Lamb \ra n (\Lam) \rho^0$ can be obtained 
directly\cite{guo1, tuan}. Taking $f_\rho=216$MeV, $V_{ub}=0.004$,
$V_{us}=0.22$, $V_{ud}=0.975$ and $a_1=0.28$ (corresponding to $N_c \sim 2$)
we obtain
\begin{eqnarray}
B(\Lamb \ra n \rho^0)=\left \{
            \begin{array}{ll}
            1.61 \times 10^{-8}& \mbox{ for $b=1.18GeV^{-1}$,}\\
            4.14 \times 10^{-9}& \mbox{ for $b=1.77GeV^{-1}$,}\\
            \end{array}
                                                 \right.     
\label{3j}
\vspace{2mm}
\end{eqnarray}
and 
\begin{eqnarray}
B(\Lamb \ra \Lam \rho^0)=\left \{
            \begin{array}{ll}
            1.23 \times 10^{-9}& \mbox{ for $b=1.18GeV^{-1}$,}\\
            3.06 \times 10^{-10}& \mbox{ for $b=1.77GeV^{-1}$.}\\
            \end{array}
                                                 \right.     
\label{3k}
\vspace{2mm}
\end{eqnarray}

In Ref.\cite{cheng1} the $\Lamb \ra n (\Lam)$ transition matrix elements
are calculated in the nonrelativistic quark model. The form factors $f_i$
and $g_i$, which are defined by ($q=p_{\Lamb} -p_f$) 
\begin{eqnarray}
 \langle f (p_f)\mid\bar{q}\gamma_{\mu}(1-\gamma_{5})b\mid
 \Lambda_{b}(p_{\Lambda_{b}})\rangle&=
&\bar{u}_{f}\{f_{1}(q^{2})\gamma_{\mu} + {i}f_{2}(q^{2})\sigma
 _{\mu\nu}q^{\nu} + f_{3}(q^{2})q_{\mu} \nn\\
&&-[g_{1}(q^{2})\gamma_{\mu} 
+ {i}g_{2}(q^{2}) 
 \sigma_{\mu\nu}q^{\nu} + g_{3}(q^{2})q_{\mu}]\gamma_{5}\}u_{\Lambda_{b}},\nn\\
&&
\label{3l}
\vspace{2mm}
\end{eqnarray}
are found to be: $f_1(0)=0.045$, $f_2(0)=-0.024/m_{\Lamb}$,
$f_3(0)=-0.011/m_{\Lamb}$,
$g_1(0)=0.095$, $g_2(0)=-0.022/m_{\Lamb}$, $g_3(0)=-0.051/m_{\Lamb}$
for $\Lamb \ra n$, and
$f_1(0)=0.062$, $f_2(0)=-0.025/m_{\Lamb}$, $f_3(0)=-0.008/m_{\Lamb}$,
$g_1(0)=0.108$, $g_2(0)=-0.014/m_{\Lamb}$, $g_3(0)=-0.043/m_{\Lamb}$
for $\Lamb \ra \Lam$.
Pole dominance behavior for the $q^2$ dependence of the form factors 
is assumed,
\begin{equation}
f_i (q^2)=\frac{f_i(0)}{\left(1-\frac{q^2}{m_{V}^{2}}\right)^2}, \;\;\;
g_i (q^2)=\frac{g_i(0)}{\left(1-\frac{q^2}{m_{A}^{2}}\right)^2}, 
\label{3m}
\vspace{2mm}
\end{equation}
where for $b \ra d$, $m_V=5.32$GeV, $m_A=5.71$GeV, and 
for $b \ra s$, $m_V=5.42$GeV, $m_A=5.86$GeV.  
Substituting Eqs.(\ref{3l}) and (\ref{3m}) into Eq.(\ref{3h}) we find that
\begin{equation}
B(\Lamb \ra n \rho^0)=6.33 \times 10^{-8},\;\;\;
B(\Lamb \ra \Lam \rho^0)=4.44 \times 10^{-9}.
\label{3n}
\vspace{2mm}
\end{equation}
These results are bigger than those in Eqs.(\ref{3j}) and (\ref{3k}). Combining
the predictions in these two models we expect that $B(\Lamb \ra n \rho^0)$ 
is around $10^{-8}$ and $B(\Lamb \ra \Lam \rho^0)$ is about $10^{-9}$.
For comparison, in $B$ decays, 
the branching ratio for $B^- \ra \pi^- \rho^0$ is
of the order $10^{-6}$\cite{bsw}, and for $B^- \ra \rho^- \rho^0$
the branching ratio is about $10^{-5}$\cite{bsw}. Hence $B(\Lamb \ra n \rho^0)$
is two to three orders smaller than those for the corresponding meson decays.

\vspace{0.2in}
{\large\bf IV. Summary and discussions}
\vspace{0.2in}

In this work we studied direct CP violation in $\Lamb$ hadronic decays 
$\Lamb\ra f \rho^0 (\omega) \rightarrow f\pi^+\pi^-$ ($f=n$ or $\Lam$).
It was found that, as a result of the inclusion of $\rho-\omega$ mixing, the 
CP-violating asymmetries in these two processes 
could be very large when the invariant
mass of the $\pi^+\pi^-$ pair is in the vicinity of the $\omega$ resonance.
For $N_c =2$, 
the maximum CP-violating asymmetries were more than 50\% and 68\% for 
$\Lamb\rightarrow n \pi^+\pi^-$ and $\Lamb\rightarrow \Lam \pi^+\pi^-$,
respectively, for reasonable values of $q^2/m_{b}^{2}$. 
Furthermore, we estimated the branching ratios for 
$\Lamb\rightarrow n (\Lam)\rho^{0}$ decays by using 
HQET and phenomenological models for the hadronic wave functions.
The results from the nonrelativistic quark 
model were also presented for comparison.
It was shown that the branching ratios are about $10^{-8}$ and $10^{-9}$
for $\Lamb \ra n \rho^0$ and $\Lamb \ra \Lam \rho^0$, respectively, which are
two or three orders smaller than those for the corresponding $B$ decays.
Since there will be more data on the heavy baryon, $\Lamb$, 
from different experimental groups in the future, 
it will be very interesting to 
look for such large CP-violating asymmetries in the experiments
in order to get a deeper understanding of the mechanism for CP violation.
On the other hand,
the smaller branching ratios for the $\Lamb$ hadronic decays
will make the measurements more difficult. Furthermore, the study of
CP violation in $\Lamb$ decays may provide insight into the
baryon asymmetry phenomena required for baryogenesis.

There are some uncertainties in our calculations. While discussing 
the CP violation in these two channels, we have to evaluate hadronic matrix
elements where nonperturbative QCD effects are involved. We have worked 
in the factorization approximation, which is expected
to be quite reliable because the $b$ quark decays are very energetic.
However, in this approach, the color-octet term is ignored. Hence $N_c$
has to be treated as an effective parameter which should be determined
by experiment. Although there are enough data to fix $N_c$ in $B$
decays as $N_c \sim 2$, the best value of $N_c$ for 
$\Lamb$ decays is not certain.
We gave the plots of the CP-violating asymmetries for $N_c=2$ and discussed
the situation for $N_c=3$. If $N_c=3$, the CP violation for 
$\Lamb\rightarrow \Lam \pi^+\pi^-$ is not large anymore. However, for
both $N_c=2$ and 3 there is large CP violation for  
$\Lamb\rightarrow n \pi^+\pi^-$. Our numerical results also depend on
$q^2/m_{b}^{2}$, but the behavior is mainly determined by $N_c$.
The $\rho-\omega$ mixing parameter,  $\tilde\Pi_{\rho\omega}$, also has some
experimental uncertainty, but this has little 
influence on our results.

While estimating the  branching ratios for $\Lamb\rightarrow n (\Lam)\rho^{0}$
we worked in the heavy quark limit. Since $m_b$ is much larger than the
QCD scale, $\Lambda_{\rm QCD}$, the $1/m_b$ corrections should be small.

\vspace{1cm}

\noindent {\bf Acknowledgment}:
\vspace{2mm}

This work was supported in part by the Australian Research Council and
the National Science Foundation of China.


\baselineskip=20pt


\newpage

\vspace{0.2in}

{\large \bf Figure Captions} \\
\vspace{0.2in}

Fig.1 The CP-violating asymmetry for $\Lamb\rightarrow n \pi^+\pi^-$ 
with $N_c=2$. The solid (dotted) line is for $q^2/m_{b}^{2}=0.3$ (0.5).\\
\vspace{0.2cm}

Fig.2 The CP-violating asymmetry for $\Lamb\rightarrow \Lam \pi^+\pi^-$ 
with $N_c=2$. The solid (dotted) line is for $q^2/m_{b}^{2}=0.3$ (0.5).\\
\vspace{0.2cm}
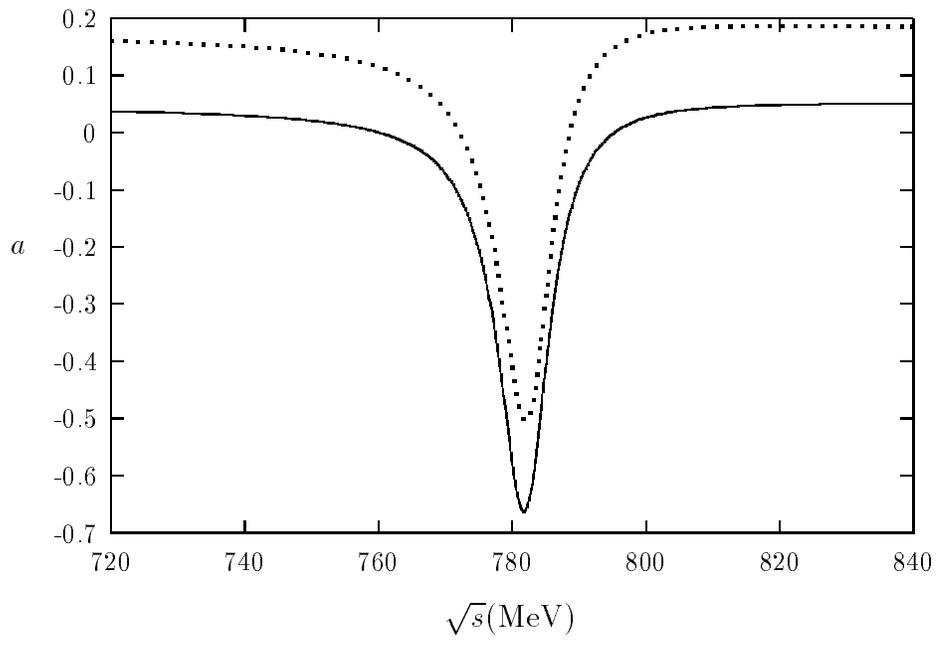
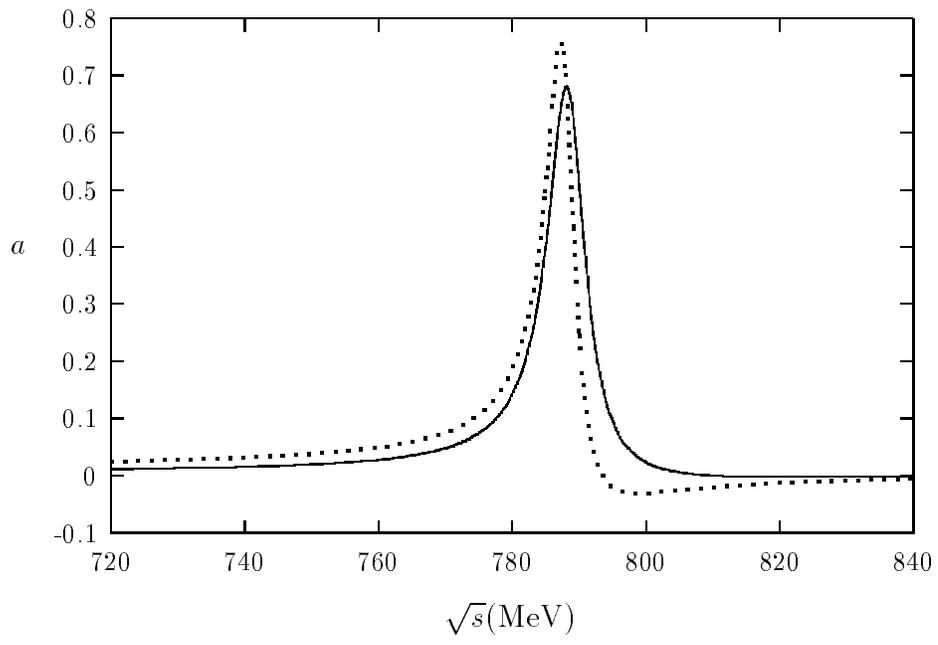
\end{document}